%% file: Romanova-lanl-new.tex
\documentclass[letterpaper,11pt,preprint]{aastex}
\usepackage{amssymb,charter,natbib}

\begin{document}

\title{Modeling of Disk-Star Interaction:
Different Regimes of Accretion and Variability}

\author{Marina M. Romanova, Akshay K. Kulkarni}
\affil{Dept. of Astronomy, Cornell University, Ithaca, NY 14853}
\email{akshay@astro.cornell.edu, romanova@astro.cornell.edu}
\author{Min Long}
\affil{Dept. of Physics, Center for Theoretical Astrophysics,
University of Illinois at Urbana-Champaign, Urbana IL, 61801}
\email{long@astro.cornell.edu}
\author{Richard V.E. Lovelace}
\affil{Dept. of Astronomy and Dept. of Applied and Eng. Physics,
Cornell University, Ithaca, NY 14853}
\email{lovelace@astro.cornell.edu}


\begin{abstract}

The appearance and time variability of  accreting millisecond
X-ray pulsars (hereafter AMXPs, e.g. Wijnands \& van der Klis 1998)
depends strongly on the accretion rate,  the effective viscosity
and the effective magnetic diffusivity of  the disk-magnetosphere boundary.
  The accretion rate is the main parameter which determines the location
of the magnetospheric radius of the star for a given stellar magnetic
field.
  We introduce a classification of
accreting neutron stars as a function of the accretion rate and show
the corresponding stages obtained from our global 3D
magnetohydrodynamic (MHD) simulations and from our axisymmetric MHD
simulations.
   We discuss the expected variability features in each
stage of accretion, both periodic and quasi-periodic (QPOs).
 We conclude that the periodicity may be
suppressed at both very high and very low accretion rates.
   In addition the periodicity may disappear when  ordered
funnel flow accretion is replaced by  disordered accretion through
the interchange instability.

\end{abstract}


\section{Different Regimes of Accretion}

\begin{figure}
\centering
\includegraphics[height=.36\textheight]{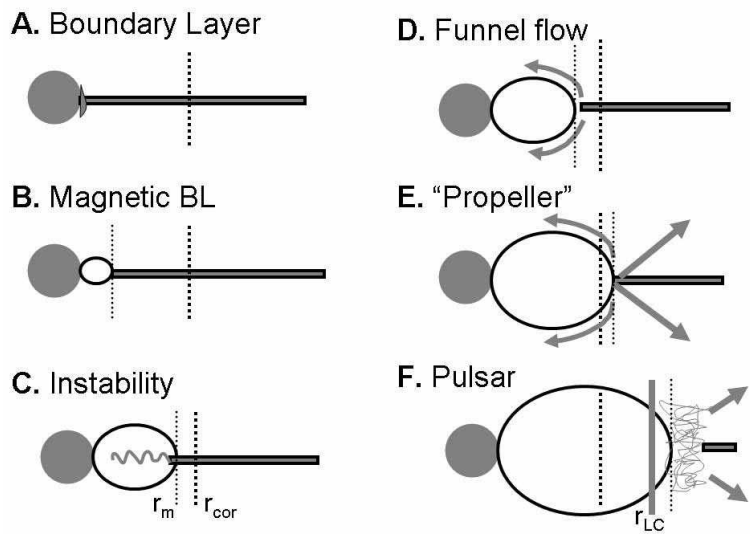}
\caption{Main regimes of accretion
 to a rotating neutron star which occur for different
  accretion rates.
   Cases A to F show different regimes when the accretion rate gradually decreases.
   A is the Boundary Layer regime where the magnetic field is completely
buried or screened, and matter accretes directly to the surface of the star;
B is the Magnetic Boundary
  Layer regime where the magnetosphere is small but may influence  the dynamics of the matter flow;
  Regimes C and D are where the magnetosphere is large enough to channel the
  matter flow;  however, accretion may also occur through instabilities
  (case C);
Regime E is the Propeller stage.  Regime F is the radio-pulsar
stage.}
\end{figure}

\noindent {\bf A. BOUNDARY LAYER (BL).} If the accretion rate is
sufficiently large  that $r_m < R_*$, then the star's magnetic field is
completely buried by the accreting matter (e.g. Cumming, Zweibel,
Bildsten 2001; Lovelace et al. 2005), and matter accretes to the
star directly through a boundary layer, which is expected to be
approximately axisymmetric because there is no channeling by the
magnetic field. In this case, high-frequency QPOs may be associated
with inhomogeneities at the boundary between inner radius of the
disk and the star. Axisymmetric hydrodynamic simulations of
accretion through the BL have been recently performed by Fisker \&
Balsara (2005).    Additional factors such as radiative
pressure in the inner regions of the disk around NS may also
determine the characteristic radius in the disk and corresponding
high-frequency QPOs (e.g., Miller, Lamb, \& Psaltis 1998; Miller \&
Lamb 2001).

\begin{figure}
\centering
  \includegraphics[height=.17\textheight]{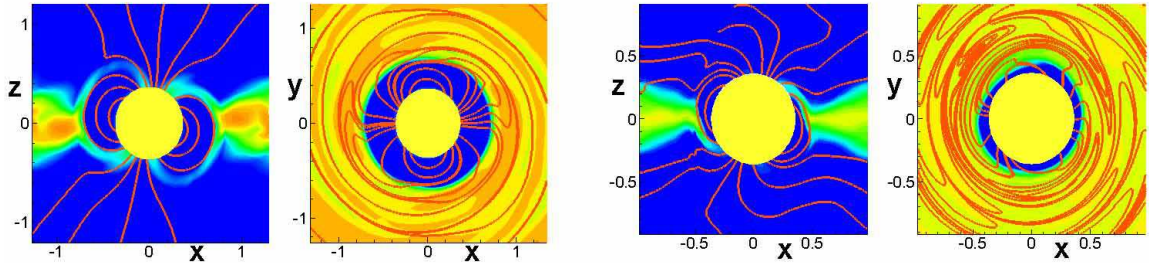}
  \caption{3D MHD simulations of accretion through
the magnetic boundary layer of a star (in the stable regime)  with a misalignment angle
  $\Theta=15^\circ$. Left two panels show slices
of density distribution and sample magnetic field lines in the case
of a small magnetosphere.
  The two right-hand panels show similar slices but for a tiny
magnetosphere.}
\end{figure}

\begin{figure*}
\centering
  \includegraphics[height=.16\textheight]{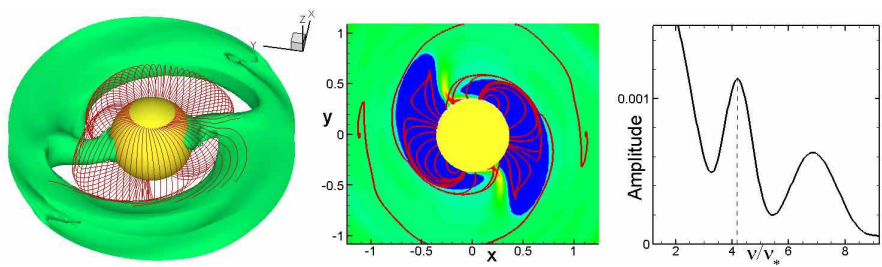}
  \caption{3D MHD simulations of accretion through an unusual magnetic boundary layer which formed in the case of a small misalignment angle $\Theta=5^\circ$.
  From left to right: 3D image of the accretion disk; an equatorial slice of the density distribution, and
  the Fourier power spectrum.}
\end{figure*}

\begin{table}
\begin{tabular}{l@{\extracolsep{0.5em}}l@{}lll}

\hline
&                                             & Type      & $\dot M$ (in $M_\odot/yr$) & Type of Variability                   \\
\hline
\multicolumn{2}{l}{$r_m < R_*$}                 & Boundary Layer          & $\dot M > 7.3\times 10^{-8}$            & QPO                     \\
\multicolumn{2}{l}{$R_* < r_m < 2 R_*$}       & Magnetic BL  & $6.5\times 10^{-9} < \dot M < 7.3\times 10^{-8}$     & QPO ($P_*$)                  \\
\multicolumn{2}{l}{$2 R_*< r_m < 3.1 R_* $}   & Funnel, Instabilities  & $1.4\times 10^{-9} < \dot M < 6.5\times 10^{-9}$   & $P_*$, QPO \\
\multicolumn{2}{l}{$3.1 R_*< r_m < 12 R_* $}  & ``Propeller"  &   $1.3\times 10^{-11} < \dot M < 1.4\times 10^{-9}$  & QPO ($P_*$)  \\
\multicolumn{2}{l}{$r_m > 12 R_* $}           & Pulsar    & $\dot M < 1.3\times 10^{-11}$    & radio $P_*$   \\
\hline
\end{tabular}
\caption{Ranges of magnetospheric radii, accretion rates and
expected variability features in different regimes of accretion for
a neutron star with $P_*=2.5 {\rm ms}$.} \label{tab:refval}
\end{table}

\medskip

\noindent{\bf B. MAGNETIC BOUNDARY LAYER (MBL)}. At lower accretion rates, the star's magnetic field is only partially buried, so that
it influences matter flow around the star. We will draw an
approximate line between MBL case and case of larger magnetospheres
at $r_m =2R_*$, so that the MBL cases correspond to $R_* < r_m < 2
R_*$. Even in this range there are a number of possible cases with
different observational properties (Romanova \& Kulkarni 2008).
\smallskip

\noindent{\it MBL with No Magnetosphere}. If the magnetospheric
radius is very close to the stellar radius, then it cannot channel
matter to the poles. However, some magnetic flux threads the inner
regions of the disk and thus may influence the dynamics of the disk
(e.g., Warner \& Woudt 2002). The authors suggested that the inner
region of the disk may form an equatorial belt, which may be
responsible for the QPOs in dwarf novae,  a sub-class of cataclysmic
variables. Such MBLs were proposed theoretically and should be
investigated in greater detail in the future. First exploratory
simulations have been done by Romanova \& Kulkarni (2008).

\smallskip

\noindent{\it  MBL with Small Magnetosphere}. We investigated
accretion to a star with a small magnetosphere in global 3D MHD
simulations (Romanova \& Kulkarni 2008). We observed from
simulations that matter may accrete in stable or unstable regime.

First we investigated accretion in stable regime to a star with
misalignment angle $\Theta=15^\circ$ in two cases of small and tiny
magnetosphere. We observed that small magnetosphere formed and
matter accreted through funnel streams (see Figure 2, left panels).
In the case of about twice as smaller dipole field of the star a
tiny magnetosphere formed and also disrupted the disk (see Figure 2,
right panels). Equatorial slice shows that the magnetic field lines
are wrapped by the disk and may influence its structure. In
particular, non-axisymmetric bending-type waves have been observed
in the inner parts of the disk.

 Another run at smaller misalignment angle $\Theta=5^\circ$ led to
{\it unstable accretion} but of unusual kind:  instead of
sporadically forming unstable tongues as in the unstable regime
(Kulkarni \& Romanova 2008a,b; Romanova, Kulkarni \& Lovelace 2008),
we observed the formation of two symmetric equatorial tongues which
connect the star with the disk and which rotate around the star with
angular velocity of the inner disk (see Figure 3). The tongues form
two equatorial spots which rotate faster than the star: each spot
rotates about 4 times faster than the star. The frequency
approximately corresponds to the frequency of the inner disk and
will change when position of the inner disk change. Thus a star in
this accretion regime will reveal high frequency QPOs with frequency
drift determined by the accretion rate to the star. Such frequency
drift and correlation with matter flux is observed in AMXPs (e.g.,
van der Klis 2000).

\begin{figure}
\centering
  \includegraphics[height=.3\textheight]{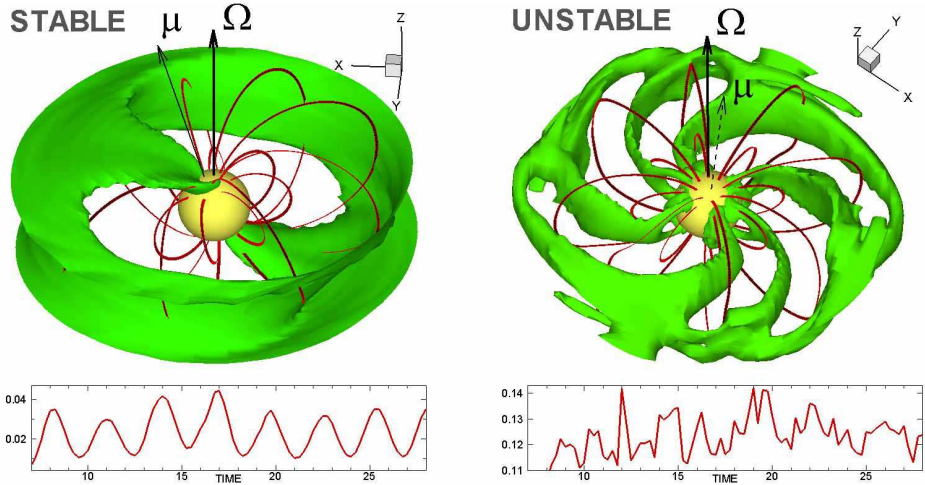}
  \caption{ Stable (left) and unstable (right) regimes of accretion.
  The bottom panels show light-curves from the hot spots
  (from Romanova, Kulkarni \& Lovelace 2008).}
\end{figure}

\medskip
\noindent{\bf C and D. UNSTABLE AND STABLE REGIMES OF ACCRETION}.
For even lower accretion rates, the disk matter is stopped by the
magnetosphere at a larger distance, a few radii of the star. Recent
3D MHD simulations have shown that in this case the matter may
accrete either through symmetric {\bf funnel streams} which are
lifted above the magnetosphere (stable regime), or through an
interchange instability, where temporary stochastic tongues of
matter form at the disk-magnetosphere boundary and penetrate deeply
into magnetosphere. These tongues form multiple hot spots on the
surface of the star. Figure 4 shows both regimes of accretion and
the corresponding light-curves associated with the hot spots on the
surface of the star.
\smallskip

\noindent {\it Stable Regime of Accretion (Funnel Streams).} In the
stable regime, matter accretes in two ordered funnel streams (Figure
4, left panel, see 3D MHD simulations by Romanova et al. 2003,
2004a; Kulkarni \& Romanova 2005). These two funnels form two
ordered spots at the surface of the star, which vary only slightly
around their preferred positions, so that the stellar period is
clearly observed in the Fourier power spectrum. This is  the regime
where X-ray pulsations in AMXPs are expected.
  If the magnetic field is more complex than the
dipole, then the hot spots have a more complex structure.
   However,  spots will have a
definite preferred locations so that a definite periodicity is
expected (Long et al. 2008).
   Our simulations  shown that a star may spin up or
down due to interaction of the magnetosphere with the disk, so that
the star's  period is expected to wander around the rotational
equilibrium state, which corresponds to the condition $r_m\approx
1.2 r_{cor}$, which has been derived from simulations by Long et al.
(2005) (see also Ghosh \& Lamb 1979). Such variations of the period
have been recently observed, e.g., by Burderi et al. (2007) and
Papitto et al. (2008). In addition to periodic variations associated
with the hot spots, QPOs associated with oscillations of the inner
disk regions are expected (Lovelace \& Romanova 2007;  Lai \& Zhang
2008).

 Simulations have shown that at small misalignment angles
$\Theta < 5^\circ$ a hot spot may rotate faster or slower than the
star depending on the accretion rate (Romanova et al. 2003; Romanova
et al. 2006). This may lead to the rapid phase variations and other
interesting phenomena observed in AMXPs (Lamb et al. 2008a,b).

\begin{figure}
\centering
  \includegraphics[height=.25\textheight]{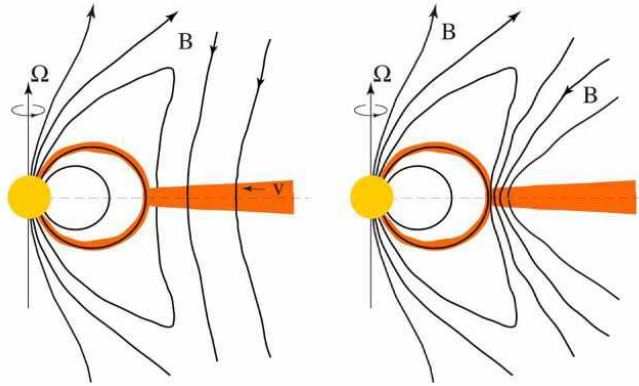}
  \caption{Disk-magnetosphere interaction for different  magnetic Prandtl numbers.
    The left-hand panel is for  $Pr=1$, when
  the disk matter flows in and penetrates through the magnetosphere diffusively at the same rate.
    The right-hand panel is for
  $Pr>>1$, where the radial accretion speed is significantly
larger than the diffusion speed of the field through the matter.}
\end{figure}

\begin{figure}
\centering
  \includegraphics[height=.2\textheight]{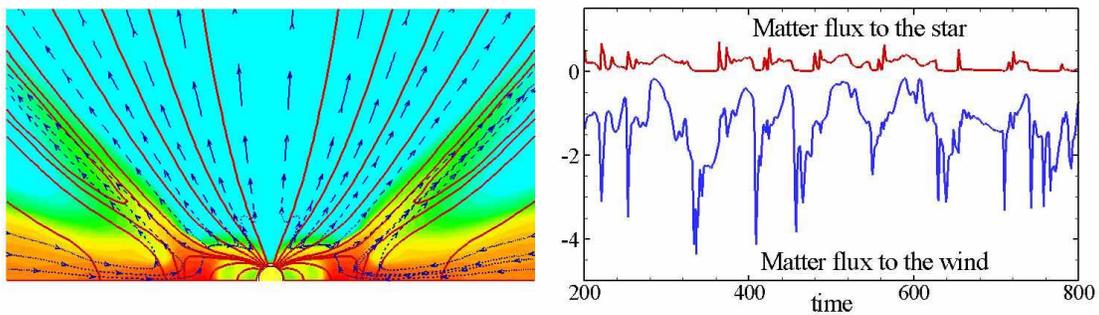}
  \caption{The left panel shows outflow of matter (background,
  selected field lines (red) and streamlines (blue) in the strong propeller
regime. The fluxes at the right panel show that the process of
accretion and outflows is episodic (from Romanova et al. 2005).}
\end{figure}

\smallskip

\noindent The {\it Unstable Regime of Accretion} has been observed
in multiple 3D MHD simulations in cases where the misalignment angle
of the dipole is small, $\Theta=5^\circ$ (Kulkarni \& Romanova
2008a,b; Romanova et al. 2008).  In the unstable regime, matter
accretes to the star through several ``tongues" which form at the
inner edge of the disk and accrete to the surface of the star on the
dynamical time-scale, forming temporary hot spots on its surface.
The light-curve from the hot spots looks stochastic (see Figure 4,
right panel), and in many cases no period is detected in the wavelet
or Fourier spectra. Thus, in the strongly unstable regime, a star
with a full-sized magnetosphere may be in the stochastic regime of
accretion. In less strongly unstable cases, both tongues and funnels
are present, and periodicity can be detected.
\smallskip

\noindent{\it  Boundary Between Stable and Unstable Regimes:}
 We observed from numerous simulations that (1) the stable
regime usually dominates at high misalignment angles, $\Theta
\gtrsim 30^\circ$; (2) enhancement of the accretion rate usually
leads to the onset of the unstable regime (Kulkarni \& Romanova,
these Proceedings). The last condition can be interpreted as
follows: for a NS with a given period of rotation $P_*$, the growth
of the accretion rate leads the gravitational force to dominate over
the centrifugal (or, $r_m << r_{cor}$),  and the instability appears
more easily. At smaller accretion rates, the magnetospheric radius
moves out, closer to $r_{cor}$, and the centrifugal force becomes
strong enough to oppose instability. Detailed comparisons of
stable/unstable regimes with instability criteria were also
performed (Kulkarni \& Romanova 2008a; Romanova et al. 2008).  From
other side we observed that at larger accretion rate the density of
matter at the inner edge of the disk is larger which is favorable
for the onset of instability.

\medskip

\noindent{\bf F. ``PROPELLER'' REGIME}. At even smaller accretion
rates, the magnetospheric radius $r_m$ becomes larger than the
corotation radius $r_{\rm cor}$, and the star enters the
``propeller" regime. Axisymmetric simulations performed in our group
have shown that propeller regime may be either $\it strong$ or $\it
weak$.

\noindent{\it In the Strong Propeller Regime}, disk matter acquires
angular momentum from the rotating magnetosphere fast enough that
most of it is ejected in conical outflows.
    At the same time a
significant amount of angular momentum and energy flows along the
open stellar field lines, giving an axially-directed Poynting-flux
dominated jet (Romanova et al. 2005; Ustyugova et al. 2006).
  Multiple simulations have shown that
the strong propeller operates when $Pr >> 1$. In this case the field
lines are bunched into an X-type point (see Figure 5, right panel)
which is favorable for outflows, as discussed by Shu and
collaborators (e.g., Cai et al. 2008). In this regime, the cycles of
matter accumulation at the inner disk, its diffusion through
magnetosphere and outbursts to jets are observed with some accretion
to the star (see also Goodson et al. 1997). Smaller ratios
$r_m/r_{cor}$ and higher diffusivities (still at $Pr \gg 1$) are
also favorable for the strong propeller regime. Oscillations are
often quasi-periodic, with frequency $\nu_{QPO}=0.01-0.2 \nu_*$,
where the lowest frequency is observed at the lowest diffusivity,
and vice-versa.
\smallskip

\noindent{\it In the Weak Propeller Regime}, when $\Pr\approx 1$,
the magnetosphere interacts with the disk, and the star loses its
angular momentum without outflows. The disk may oscillate due to
such interaction, and most of the incoming disk matter may accrete
to the star's surface (Romanova et al. 2004b). It is often the case
that the rotating magnetosphere (modified by the azimuthal wrapping
of the field) pushes the disk to larger distances, and matter may
accrete from that point. In the limit of  a very weak propeller
where $r_m\approx r_{cor}$, stellar pulsations and different QPO
frequencies are expected as in the case of funnel flow accretion. If
magnetosphere is large enough then accretion may be stopped by the
centrifugal barrier of the fast rotating magnetosphere in both
``propeller" regimes.

\medskip

\noindent{\bf G. PULSAR REGIME}. ~For even smaller accretion rates,
the radius of the magnetosphere becomes larger than the light
cylinder radius, $r_{L}=c/\Omega_*$ . In this regime, accretion to
the surface of the NS is suppressed, and the star becomes a
radio-pulsar. This stage is least investigated numerically and it is
not clear, for example, at what values of $r_m/r_{LC}$ the
radio-pulsar emission dominates.

\section{Discussion}

 Analysis of different stages of
evolution of accreting magnetized neutron stars have shown that they
have different appearances for different accretion rates. In
addition, their appearance may be different depending on other
factors, such as the magnetic diffusivity and the Prandtl number
(ratio of viscosity to diffusivity).
    We chose as an example a typical NS with period
$P_*=2.5~{\rm ms}$ ($\nu_*=400 {\rm Hz})$, and derived accretion
rates from equation (2) corresponding to the boundaries between
different regimes.
   The results are shown in the Table. As we see from
simulations of different  regimes, periodic oscillations (with the
star's period) are observed only in the regime of magnetospheric
accretion which corresponds to a rather narrow range of accretion
rates, $1.4\times 10^{-9}~{M_\odot/yr} < \dot M < 6.5\times
10^{-9}{M_\odot/yr}$. This fact may explain the appearance of
pulsations in intermittent AMXPs in a narrow range of $\dot M$ as
discussed by Altamirano et al. (2008) in the case of SAX
J1748.9-2021. If the pulsations are also present in the magnetic
boundary layer case and in the weak propeller regime, then these
boundaries for  $\dot M$ will be somewhat wider.

\bibliography{ms}

\end{document}